\documentclass[aps,prd,reprint,twocolumn,superscriptaddress,showpacs]{revtex4-1}
\usepackage{CJK}
\usepackage{graphicx}
\usepackage{mathrsfs}
\usepackage{bm}
\usepackage{amsmath}
\usepackage{dcolumn}
\usepackage{epstopdf}
\usepackage{dsfont}
\usepackage{amssymb}
\usepackage{tabularx}
\usepackage{array}
\usepackage{float}
\usepackage{color}
\usepackage{epstopdf}
\usepackage{mathrsfs}
\usepackage[colorlinks, linkcolor=blue,anchorcolor=blue,citecolor=blue,urlcolor=blue]{hyperref}

\begin{document}

\title{Quadratic contact point semimetal: Theory and material realization}

\author{Ziming Zhu}
\affiliation{Research Laboratory for Quantum Materials, Singapore University of Technology and Design, Singapore 487372, Singapore}
\affiliation{Key Laboratory of Low-Dimensional Quantum Structures and Quantum Control of Ministry of Education, Department of Physics and Synergetic Innovation Center for Quantum Effects and Applications, Hunan Normal University, Changsha 410081, China}

\author{Ying Liu}\email{ying\_liu@mymail.sutd.edu.sg}
\affiliation{Research Laboratory for Quantum Materials, Singapore University of Technology and Design, Singapore 487372, Singapore}

\author{Zhi-Ming Yu}\email{zhiming\_yu@sutd.edu.sg}
\affiliation{Research Laboratory for Quantum Materials, Singapore University of Technology and Design, Singapore 487372, Singapore}

\author{Shan-Shan Wang}
\affiliation{Research Laboratory for Quantum Materials, Singapore University of Technology and Design, Singapore 487372, Singapore}

\author{Y. X. Zhao}
\affiliation{National Laboratory of Solid State Microstructures and Department of Physics, Nanjing University, Nanjing 210093, China}
\affiliation{Collaborative Innovation Center of Advanced Microstructures, Nanjing University, Nanjing 210093, China}

\author{Yuanping Feng}
\affiliation{Department of Physics, National University of Singapore, Singapore 117542, Singapore}

\author{Xian-Lei Sheng}\email{xlsheng@buaa.edu.cn}
\address{Department of Physics, Key Laboratory of Micro-nano Measurement-Manipulation and Physics (Ministry of Education), Beihang University, Beijing 100191, China}

\author{Shengyuan A. Yang}
\affiliation{Research Laboratory for Quantum Materials, Singapore University of Technology and Design, Singapore 487372, Singapore}
\affiliation{Center for Quantum Transport and Thermal Energy Science, School of Physics and Technology, Nanjing Normal University, Nanjing 210023, China}

\begin{abstract}
Most electronic properties of metals are determined solely by the low-energy states around the Fermi level, and for topological metals/semimetals, these low-energy states become distinct because of their unusual energy dispersion and emergent pseudospin degree of freedom. Here,
we propose a class of materials which are termed as quadratic contact point (QCP) semimetals. In these materials, the conduction and valence bands contact at isolated points in the Brillouin zone, around which the band dispersions are quadratic along all three directions. We show that in the absence/presence of spin-orbit coupling, there may exist triply-/quadruply-degenerate QCPs that are protected by the crystalline symmetry. We construct effective models to characterize the low-energy fermions near these QCPs. Under strong magnetic field, unlike the usual 3D electron gas, there appear unconventional features in the Landau spectrum. The QCP semimetal phase is adjacent to a variety of topological phases. For example, by breaking symmetries via Zeeman field or lattice strain, it can be transformed into a Weyl semimetal with Weyl and double Weyl points, a $\mathbb{Z}_2$ topological insulator/metal, or a Dirac semimetal. Via first-principles calculations, we identify realistic materials Cu$_2$Se and RhAs$_3$ as candidates for QCP semimetals.
\end{abstract}

\maketitle

\section{Introduction}

In the past few years, topological metals/semimetals have emerged as an intriguing area of research in condensed matter physics~\cite{Chiu_RMP,Bansil_RMP,burkov2016topological,yang2016sa,dai2016quantum,Armitage_RMP}. In these materials, the electronic band structures feature topology/symmetry-protected band degeneracies near the Fermi level, such that the low-energy fermionic excitations around the band-degenerate manifold behave distinctly from the conventional Schr\"{o}dinger-type fermions. As most electronic properties for metals are determined by the states near the Fermi surface, this also means that these materials would exhibit highly nontrivial physical properties, e.g., in transport, magnetic, and optical responses~\cite{nielsen1983adler,yang2011quantum,son2013chiral,jian2013topological,burkov2014chiral,xiong2015evidence,huang2015observation,shekhar2015extremely,zhang2016linear,gao2017intrinsic,liu2018circular}.

For a three-dimensional (3D) material, the band-degenerate manifold may take the form of a 0D point, a 1D line, or even a 2D surface in the Brillouin zone (BZ). Of these possibilities, the point band-degeneracy is perhaps the mostly studied so far. For example, the Weyl semimetals are materials with
two-fold degenerate linear band-crossings points, around which the low-energy electrons resemble the Weyl fermions long sought after in high-energy physics~\cite{murakami2007phase,wan2011topological,burkov2011weyl}. A point band-degeneracy is characterized in at least the following \emph{three aspects}. The first is the band dispersion around the point. For example, the dispersion is linear along all three directions for a Weyl point, whereas for so-called double Weyl point, the dispersion is quadratic in two directions and linear in the remaining one~\cite{xu2011chern,Fang2012,huang2016new}. The second is the degeneracy of the point~\cite{bradlyn2016beyond}. For example, a Dirac point is four-fold degenerate, such that the low-energy states necessarily carry an inherent four-component spinor form~\cite{young2012dirac,A3Bi_wang,CdAs_wang,liu2014discovery,liu2014stable}, in contrast to the Weyl points that are two-fold degenerate.
The third aspect is regarding the stability of the degenerate point, i.e., the mechanism that protects the point needs to be clarified. For example, a Weyl point is topologically protected by the quantized Chern invariant $\pm 1$ for a constant energy surface enclosing the point~\cite{wan2011topological,Zhao2013b}, whereas a Dirac point would require crystalline symmetries for protection~\cite{A3Bi_wang}.

The first and the second aspects above determine the most essential features of the low-energy fermions: the energy dispersion and the internal (pseudospin) degree of freedom. The third aspect determines possible topological phase transitions under symmetry breaking. A current theme of research is to explore novel types of band-degeneracies that is new in at least one of the three aspects. Meanwhile, equally important is the task of searching for realistic material systems for the realization of such new topological phases.

In this work, we theoretically propose a class of materials that may be termed as quadratic contact point (QCP) semimetals. In a QCP semimetal, the
conduction and valence bands contact at isolated points, where the band dispersion is quadratic along \emph{all} three directions. We find that in the absence/presence of spin-orbit coupling (SOC), the QCP can be triply-/quadruply-degenerate, and the stability of the point can be protected by symmorphic crystalline symmetries. It needs to be emphasized that although the quadratic dispersion is similar to that for the electron gas model,
a key difference of QCP fermions is its intrinsic multi-component (pseudospin) degree of freedom, which leads to distinct and much richer physics. This is reflected in the effective models and in the unconventional Landau level (LL) spectrum. We further show that the QCP semimetal phase may be regarded as a parent phase for a variety of topological phases, including the Weyl semimetal with Weyl and double Weyl points, the $\mathbb{Z}_2$ topological insulator/metal, and the Dirac semimetal. The topological phase transitions can be achieved by proper symmetry breaking, e.g. by Zeeman field or lattice strain. By using first-principles calculations, we identify realistic materials as possible candidates for QCP semimetals. Our work reveals interesting fundamental physics for a new semimetal phase, and offers a promising platform for exploring new fermionic quasiparticles and controlled topological phase transitions.

\section{QCP in the absence or presence of SOC: Concrete example}

\begin{figure}[!htp]
{\includegraphics[clip,width=8.2cm]{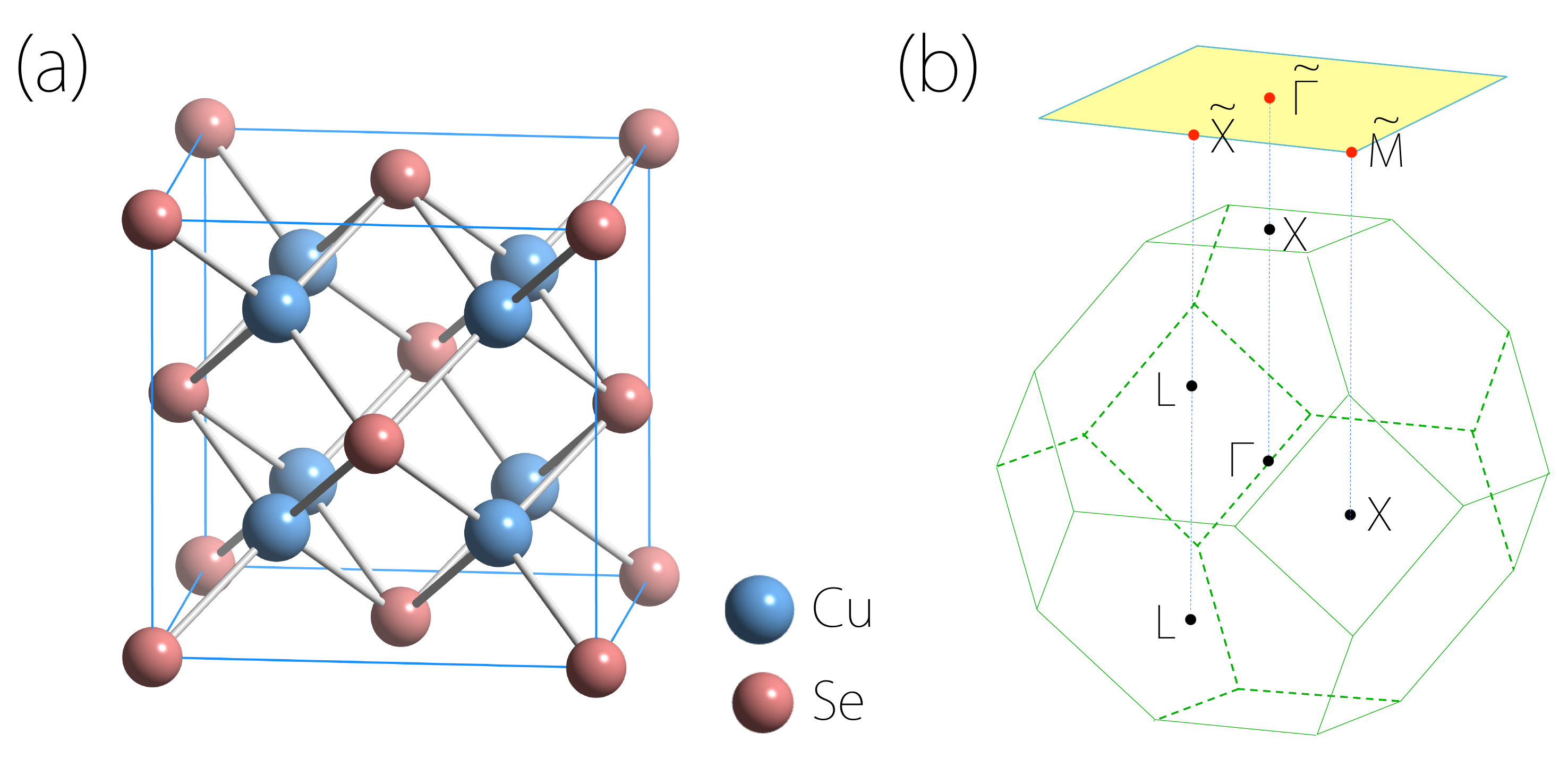}}
\caption{\label{fig:Cu2Se}
(a) Conventional unit cell for ${\rm Cu_{2}Se}$,  and (b) the corresponding bulk and surface Brillouin zones.}
\end{figure}

We motivate our discussion by investigating the band structure for a concrete example material first. We consider ${\rm Cu_{2}Se}$ , which takes the antifluorite structure with space group No.~225 ($Fm\bar{3}m$). The material has been synthesized in experiment~\cite{yamamoto1991x}.  As shown in Fig.~\ref{fig:Cu2Se}(a), the conventional unit cell has a cubic shape, and the structure can be regarded as a double-nested zinc-blende lattice. The experimentally measured lattice constant is $a=4.075$ \AA~\cite{yamamoto1991x}, which is used in our calculation.

To study the band structure, we perform the first-principles calculations based on the density functional theory (DFT). The calculation details are presented in Appendix~\ref{sec:meth}. Figure~\ref{fig:Cu2Se_noSOC} shows the result in the absence of SOC. From the projected density of states (PDOS) plot (right panel of Fig.~\ref{fig:Cu2Se_noSOC}), one observes that the system is a semimetal with suppressed DOS at the Fermi level. The low-energy bands are mainly contributed by the $s$ and $d$ orbitals of Cu atom and $p$ orbital of Se atom. Importantly, one finds that the conduction and the valence bands contact at a single Fermi point at $\Gamma$. Here, while the conduction band is non-degenerate (without counting spin), the valence band top is actually doubly degenerate. From Fig.~\ref{fig:Cu2Se_noSOC}, one observes that the two valence bands are degenerate along the high-symmetry paths $\Gamma$-L and $\Gamma$-X, but they split along other directions such as $\Gamma$-W. Moreover, around the band contact point, the bands have quadratic dispersion along all directions. Thus, this point qualifies as a QCP, and it is a triply degenerate point (if counting spin, it will become sixfold-degenerate).

\begin{figure}[!htp]
{\includegraphics[clip,width=6.4cm]{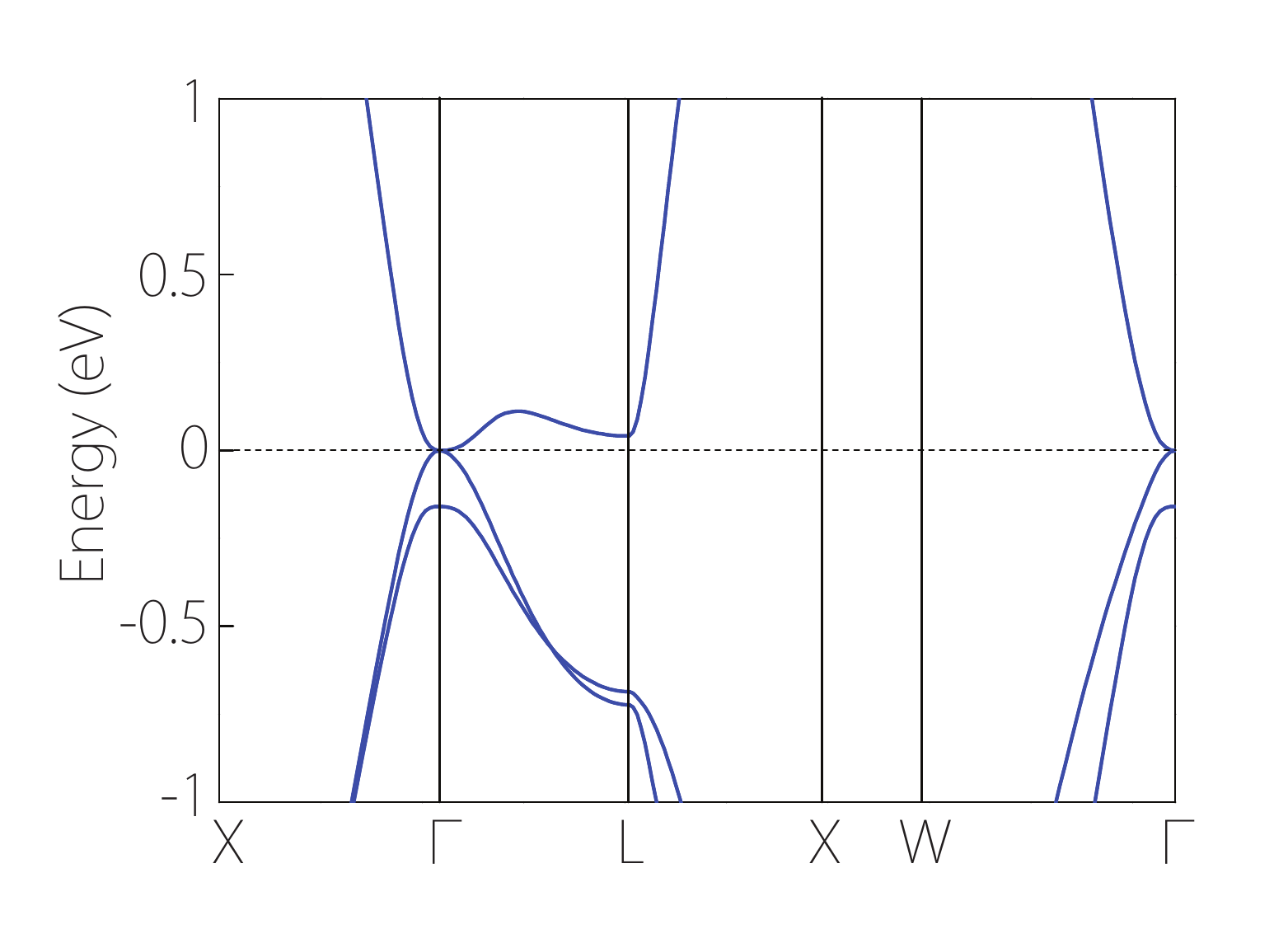}}
\caption{\label{fig:Cu2Se_SOC}
Calculated band structure for bulk ${\rm Cu_{2}Se}$ with SOC included.}
\end{figure}

Figure~\ref{fig:Cu2Se_SOC} shows the DFT band structure when SOC is included. Due to the presence of inversion symmetry ($\mathcal{P}$) and time reversal symmetry ($\mathcal{T}$), each band is at least doubly (spin-)degenerate. The original degenerate valence band top is split by SOC. However, the upper valence band still touches the conduction band at $\Gamma$, around which both bands have quadratic dispersions. Hence this band contact point is still a QCP. Here, in the presence of SOC, spin must be explicitly counted, so this QCP has four-fold degeneracy.

From the above discussion, we find that: for the material ${\rm Cu_{2}Se}$, it is a semimetal with a triply-degenerate QCP in the absence of SOC, and it has a quadruply-degenerate QCP in the presence of SOC.

We have two remarks before proceeding. First, one might think that the quadratically dispersing bands around the QCP are not very different from the usual single-band electron gas model for metals. This is not the case. A salient difference is that the low-energy states near a QCP necessarily have an internal multi-component pseudospin degree of freedom. This will in turn give rise to distinct features in the LL spectrum, as we shall discuss in the following sections. Second, it is known that quadratic band touching point also appears between the (heavy hole and light hole) valence bands for certain semiconductors like GaAs. In contrast to that case, the QCP here is between conduction and valence bands, which dictates a gapless system. We shall see that this point is also crucial for the QCP semimetal to be a parent phase for other topological semimetal/insualtor phases, which does not happen for the conventional semiconductors such as GaAs.

\begin{figure}[!htp]
{\includegraphics[clip,width=8.4cm]{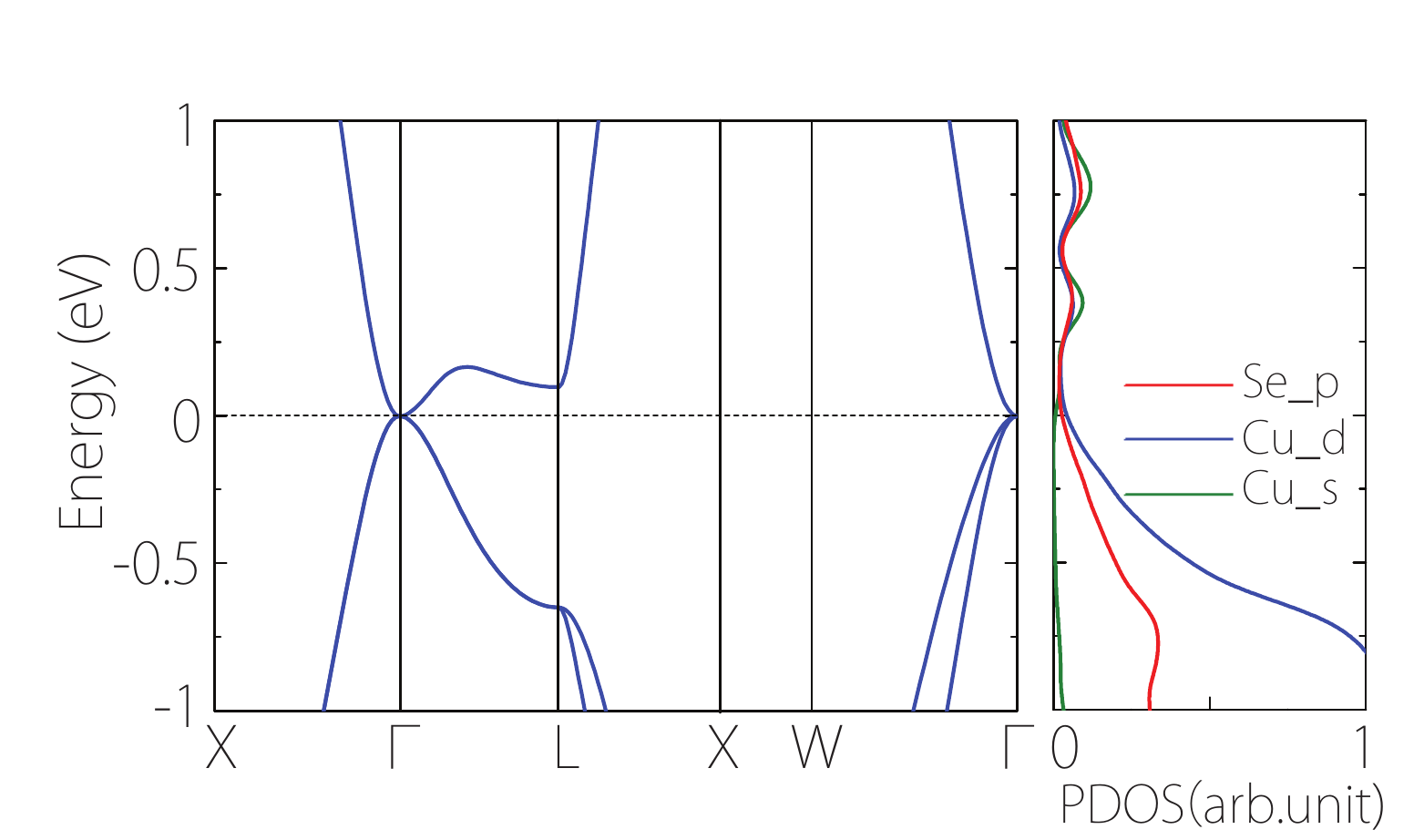}}
\caption{\label{fig:Cu2Se_noSOC}
Calculated band structure for bulk ${\rm Cu_{2}Se}$ without SOC. The right panel shows the projected density of states (PDOS).}
\end{figure}

\section{Symmetry protection and effective models}

Before our analysis, we should point out that although the discussion so far is with the concrete material example (${\rm Cu_{2}Se}$) in Sec.~II, the analysis for the QCPs below is quite general, depending only on the symmetry of the system and not limited to the particular example material. Hence the resulting models can be directly applied to other material systems with the same symmetry.

Let us first consider the band structure in Fig.~\ref{fig:Cu2Se_noSOC} in the absence of SOC. Here, the two valence bands are degenerate along the $\Gamma$-L and $\Gamma$-X paths. The QCP at the $\Gamma$ point is associated with the point group symmetry of the crystal, which is the $O_h$ cubic point group. The QCP is protected because the three degenerate states there belong to the three-dimensional $T_{2g}$ irreducible representation. The generators for the $O_h$ group may be taken as: {$C_{4z}$, $M_{[110]}$, $C_{3,[111]}$ and $\mathcal{P}$}. These symmetry operations constrain the form of the effective model, as in the standard $k\cdot p$ approach.

In the basis of the three degenerate states, we find that the effective model expanded around $\Gamma$ takes the following form (up to quadratic order in $k$, and we set $\hbar=c=1$ hereafter)

\begin{equation}\begin{split}\label{H1}
&\mathcal{H}_1(\bm k)= Ak^2+C(k_xk_y \lambda_1-k_x k_z \lambda_4+k_y k_z\lambda_6)\\
& +D\ \text{diag}(2k_x^2-k_y^2-k_z^2, 2k_y^2-k_x^2-k_z^2, 2k_z^2-k_x^2-k_y^2).
\end{split}
\end{equation}
Here, the energy and the wave-vector $\bm k$ are measured from the QCP, $\text{diag}$ stands for a diagonal matrix, $\lambda_i$ ($i=1,4,6$) are the three symmetric Gell-Mann matrices:
\begin{eqnarray}
\lambda_{1}=\left[\begin{array}{ccc}
0 & 1 & 0\\
1 & 0 & 0\\
0 & 0 & 0
\end{array}\right],~& \lambda_{4}=\left[\begin{array}{ccc}
0 & 0 & 1\\
0 & 0 & 0\\
1 & 0 & 0
\end{array}\right], ~& \lambda_{6}=\left[\begin{array}{ccc}
0 & 0 & 0\\
0 & 0 & 1\\
0 & 1 & 0
\end{array}\right], \nonumber\\
\end{eqnarray}
and $A$, $C$, $D$ are material-specific real valued parameters. The form of the effective model clearly indicates that the band dispersion is quadratic around the QCP. Along a momentum axis, say $k_x$, the three bands have dispersions:
\begin{equation}
\epsilon_{1,2}=(A-D)k_x^2,\qquad \epsilon_3=(A+2D)k_x^2.
\end{equation}
Hence, to make the degeneracy point at $k=0$ a QCP, i.e., a contact point between conduction and valence bands (the bands bend in opposite ways in energy), a necessary condition
\begin{equation}
(A-D)(A+2D)<0
\end{equation}
must be satisfied. For the band structure of Cu$_{2}$Se in Fig.~\ref{fig:Cu2Se_noSOC}, we should have $(A-D)<0$ and $(A+2D)>0$.
Indeed, by fitting the band structure using the model (\ref{H1}),
we obtain that $A=-36.02$ eV$\cdot${\AA}$^{2}$, $D=131.11$ eV$\cdot${\AA}$^{2}$, and $C=-122.71$ eV$\cdot${\AA}$^{2}$, which satisfy the condition that we have mentioned.

When SOC is included, we need to consider the double representations for the corresponding symmetry group. By analyzing the symmetry for the degenerate states at the fourfold QCP in Fig.~\ref{fig:Cu2Se_SOC}, we find that they belong to the four-dimensional $\Gamma_8^-$ irreducible representations. The basis states for this representation can be chosen as the $J=3/2$ multiplet (with certain symmetry-enforcing factor), based on which the symmetry-constrained effective model expanded around the QCP can be written in the following form (again up to second order in $k$)
\begin{eqnarray}\label{H2}
\mathcal{H}_2(\bm k)&=&\alpha k^{2}+\frac{\sqrt{3}}{2}\beta(k_{x}^{2}-k_{y}^{2})\Gamma_1 +\frac{\beta}{2}(2k_{z}^{2}-k_{x}^{2}-k_{y}^{2})\Gamma_2 \nonumber\\
&&+\gamma(k_{x}k_{y}\Gamma_3+k_{x}k_{z}\Gamma_4+k_{y}k_{z}\Gamma_5),
\end{eqnarray}
where $\Gamma_1=\sigma_x\otimes \sigma_0$, $\Gamma_2=\sigma_z\otimes \sigma_z$, $\Gamma_3=\sigma_y\otimes \sigma_0$, $\Gamma_4=\sigma_z\otimes \sigma_x$, and $\Gamma_5=\sigma_z\otimes \sigma_y$ are the five $4\times 4$ $\Gamma$ matrices realizing the Clifford algebra $\{\Gamma_a,\Gamma_b\}=2\delta_{ab}$, $\sigma_0$ is the $2\times 2$ identity matrix, $\sigma_{x,y,z}$ are the Pauli matrices, and $\alpha$, $\beta$, $\gamma$ are material-specific real valued model parameters. One notes that besides the diagonal term $\alpha k^2$, the remaining terms all anti-commute with each other. It follows that the four bands are pairwise degenerate (consistent with the $\mathcal{PT}$ symmetry), with the dispersion
\begin{equation}\label{disp}
\epsilon=\alpha k^2\pm \sqrt{\beta^2\sum_{i}k_i^4 +\frac{1}{2}(\gamma^2-\beta^2)\sum_{i\neq j}k_i^2 k_j^2},
\end{equation}
where the subscripts $i,j\in\{x,y,z\}$.
In order for this model to describe a QCP, we need to further require that the second term in Eq.~(\ref{disp}) dominates the first term. In particular, for the dispersion along the $k_i$ axis, we should have the condition $|\beta|>|\alpha|$ satisfied. Using this model to fit the band structure in Fig.~\ref{fig:Cu2Se_SOC}, we obtain that $\alpha=-5.62$ eV$\cdot${\AA}$^{2}$, $\beta=179.91$ eV$\cdot${\AA}$^{2}$, and $\gamma=14.07$ eV$\cdot${\AA}$^{2}$.

The models (\ref{H1}) and (\ref{H2}) developed here describe the low-energy electronic states around the triply-degenerate and quadruply-degenerate QCPs, respectively. It is clear that these states have an emergent pseudospin degree of freedom due to the entanglement of multiple bands at the QCP.
This is in sharp contrast to the usual electron gas model, although it also has a quadratic dispersion. The difference is also reflected in the LL spectrum under a strong magnetic field, as we shall discuss in the following section.

\section{Landau spectrum for QCP fermions}

\begin{figure}
  \includegraphics[width=8.4 cm]{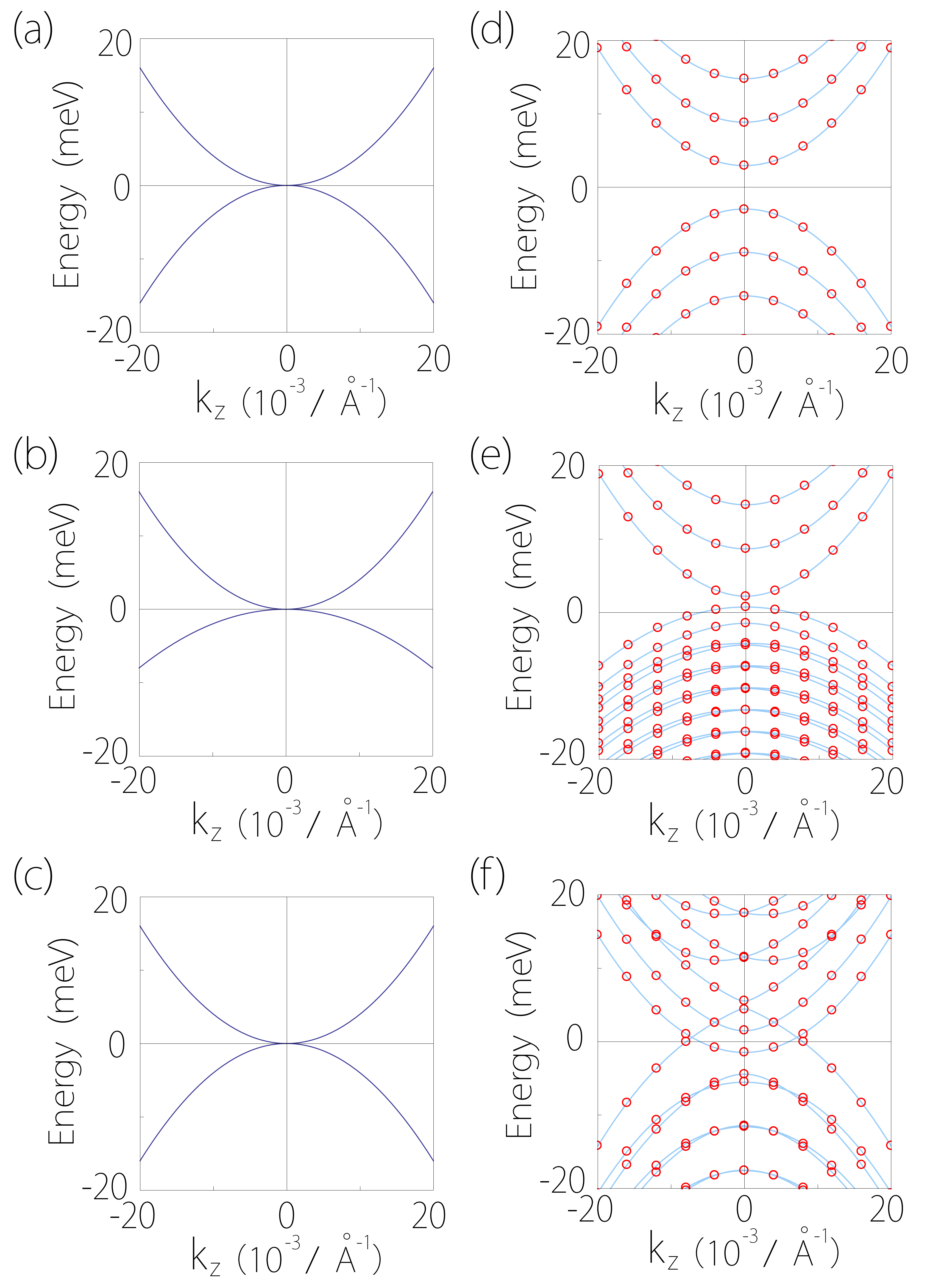}
  \caption{(a-c) Energy spectrum (along $k_z$) for (a) zero-gap semiconductor model, (b) three-component QCP model, and (c) four-component QCP model. (d-f) show the corresponding Landau spectrum for magnetic field along the $z$ direction. In the calculation, we take $B=5$ T, $m=0.1 m_e$ with $m_e$ the free electron mass for (a,d); $A=0$, $C=-60$ $\mathrm{eV}\cdot\mathrm{\AA}^2$, and {{$D=20$ $\mathrm{eV}\cdot\mathrm{\AA}^2$}} for (b,e); $\alpha=0$, $\beta=40$ $\mathrm{eV}\cdot\mathrm{\AA}^2$, and $\gamma=69$ $\mathrm{eV}\cdot\mathrm{\AA}^2$ for (c,f).}\label{LLs}
\end{figure}

The quadratic dispersion for each branch of the QCP fermion is similar to that for the usual 3D electron gas. To better visualize their difference in the Landau spectrum, we compare the results for QCP fermions to that for a conventional ``zero-gap semiconductor" model, given by
\begin{eqnarray}\label{EG}
  \mathcal H_\mathrm{ZS}=\frac{k^2}{2m}\sigma_z.
\end{eqnarray}
Here, $m$ is the effective mass, the two branches are decoupled, representing the usual electron (hole) band, {and the subscript ZS standards for the zero-gap semiconductor model}. The corresponding band structure along the $k_z$ direction is shown in Fig.~\ref{LLs}(a). We let the two bands touch at the $\Gamma$ point, such that the
band structure would look similar to those for QCP fermions as in Fig.~\ref{LLs}(b) and (c).

Despite their apparent similarity, in the following, we show that their spectra in the presence of a magnetic field will exhibit distinct differences. We take a magnetic field along the $z$ direction and focus on the orbital effect from the $B$ field. In the effective models, the $B$ field enters through the standard Peierls substitution $\bm k\rightarrow \bm k+e\bm A$, where the vector potential $\bm A=(-By,0,0)$ by using the Landau gauge. This applied magnetic field quantizes the motion of the electrons in the $x$-$y$ plane but leaves the motion in $z$ unaffected. Hence, the resulting spectrum consists of Landau subbands which disperse along $k_z$.

The result for the conventional zero-gap semiconductor model (\ref{EG}) is simple. Since the two bands are decoupled, the spectrum for each band is just like that for a 3D electron (hole) gas. The overall spectrum can be written as
\begin{equation}
E_n^\text{ZS}(B,k_z)=\pm \Big[\Big(n+\frac{1}{2}\Big)+\frac{1}{2}\ell_B^2k_z^2\Big] \omega_c,
\end{equation}
where $n=0,1,2,\cdots$, $\ell_B=\sqrt{1/eB}$ is the magnetic length, and $\omega_c=eB/m$ is the cyclotron frequency. The spectrum is plotted in Fig.~\ref{LLs}(d). One observes that bottom (top) of the electron-like (hole-like) Landau subband is located at $\pm\omega_c/2$, such that the spectrum has an energy gap of $\omega_c$.

For the QCP fermions, closed analytic forms for the Landau spectra are difficult to obtain, so we calculate the spectra numerically. And for simplicity, we drop the overall shift terms in model (\ref{H1}) and (\ref{H2}) by setting $A=0$ and $\alpha=0$. Figure~\ref{LLs}(e) shows the result for the three-component QCP fermion. One observes that the number of hole-like subbands are doubled due to the two valence bands in model (\ref{H1}), and the original double degeneracy is lifted [compare Fig.~\ref{LLs}(b) and Fig.~\ref{LLs}(e)]. Comparing Fig.~\ref{LLs}(d) and Fig.~\ref{LLs}(e), one also observes that although their dispersions without $B$ field look similar, the sizes of energy gaps induced by $B$ field can be quite different. For the special case with {{$D=-C/3$}} in (\ref{H1}), we can obtain an analytic expression for the energy gap in the Landau spectrum observed in Fig.~\ref{LLs}(e), given by {{$D/\ell_B^2$}}. {We mention that here the parameters $C$ and $D$ need to have opposite signs. Otherwise, the middle energy bands would take a hyperbolic type dispersion and must cross the Fermi level. For such cases, the model (\ref{H1}) no longer describes a QCP semimetal state.}

The situation is even more interesting for the four-component QCP fermion in Eq.~(\ref{H2}). Its Landau spectrum is shown in Fig.~\ref{LLs}(f). First of all, one finds that in contrast to the zero-gap semiconductor model, the result here does not have an energy gap in the spectrum. The top of the hole-like Landau subbands is higher than the bottom of the electron-like ones. Second, the original double degeneracy for the bands is lifted by the applied magnetic field. The Landau subbands in Fig.~\ref{LLs}(f) are nondegenerate. Third,
one notes that although the original band structure is symmetric in energy (when $\alpha=0$) [see Fig.~\ref{LLs}(c)], the Landau spectrum under $B$ field becomes asymmetric. This behavior is also different from the zero-gap semiconductor model.

\section{Topological phase transition}

From the previous discussion, we see that the QCPs are protected by the crystalline symmetries. In the example of Cu$_{2}$Se, it is the $O_h$ point group symmetry that offers the protection. The QCP point becomes unstable when the symmetry is broken, and this may lead to quantum phase transitions into other topological phases. Here, we consider two possible mechanisms for driving the phase transition: the Zeeman field and the lattice strain.

\subsection{Effect of Zeeman field}

A Zeeman field couples with the spin magnetic moment of quasiparticles. Physically, it may be realized by ordered magnetic dopants, or by external magnetic field in the regime where the orbital quantization can be neglected (typically requiring a large $g$-factor for the material and at low field). In the following, we study the effect of Zeeman field on the QCP described by the effective model in Eq.~(\ref{H2}). [Since spin needs to be considered, it is not suitable to consider the effect in model (\ref{H1}).]

In model (\ref{H2}), as we discussed in Sec.~III, the basis states transform as the $J=3/2$ multiplet, so the Zeeman term takes the form of
\begin{equation}
H_\text{Z}=-\bm{M}\cdot\bm J,
\end{equation}
where $\bm M$ represents the Zeeman field (its magnitude is the coupling strength, which includes the field strength, $g$-factor, magnetic moment, and etc.), and $\bm J$ is the vector of the $J=3/2$ spin matrices.

\begin{figure}
  \includegraphics[width=8.4 cm]{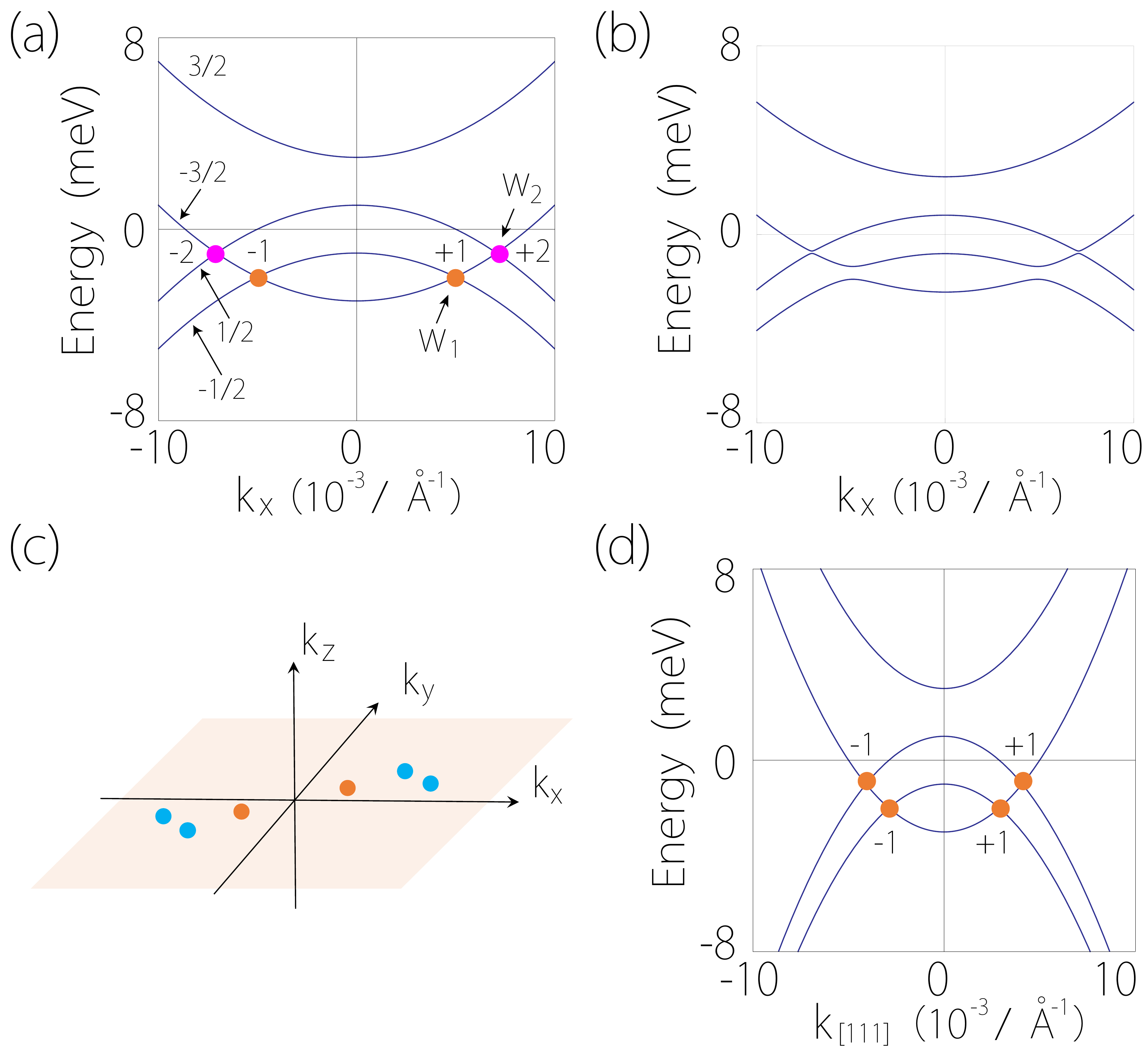}\\
  \caption{Transformation of QCP point under Zeeman field. (a) Band structure for Zeeman field along the $[100]$ direction. Orange (purple) dots stand for the single (double) Weyl points. (b) The corresponding band structure along $k_x$ when a perturbation $\delta H$ that breaks $c_{4x}$ symmetry is added. The original double Weyl points are split by the perturbation, as schematically shown in (c). Here, each original double Weyl point split into a pair of single Weyl points (blue dots), whereas the original single Weyl points (orange dots) are preserved but slightly displaced from the $k_x$ axis. (d) Band structure for Zeeman field along the $[111]$ direction. Four single Weyl points are observed. In the calculation, we take $M=2$ meV, $\alpha=0$, $\beta=40$ $\mathrm{eV}\cdot\mathrm{\AA}^2$, $\gamma=69$ $\mathrm{eV}\cdot\mathrm{\AA}^2$, $\delta_M=0.4$ meV. }\label{Zeeman}
\end{figure}

The Zeeman field breaks the time reversal symmetry, hence the original double degeneracy for each band in model (\ref{H2}) is lifted.
In the following, we show that as a result, there appear Weyl and double Weyl points in the band structure.
Let's first consider the Zeeman field along the $[100]$ direction. Then the Zeeman coupling term takes the form of $H_\text{Z}=-MJ_x$.
For simplicity, we neglect the overall shift term by putting $\alpha=0$,  because it does not affect the crossing between bands.
We first inspect the band structure along the $k_x$-axis, which is governed by the reduced Hamiltonian
\begin{eqnarray}
  \widetilde{\mathcal H}(k_x)=\frac{\sqrt{3}}{2}\beta k_x^2\Gamma_1-\frac{\beta}{2}k_x^2\Gamma_2-MJ_x.
\end{eqnarray}
The four bands on the axis are split in energy, given by
\begin{eqnarray}
  E_{J_x=\pm\frac{1}{2}}&=&-\beta k_x^2\pm\frac{1}{2}M,\nonumber\\
  E_{J_x=\pm\frac{3}{2}}&=&+\beta k_x^2\pm\frac{3}{2}M.
\end{eqnarray}
Particularly, the fourfold degeneracy at the $\Gamma$ point is completely lifted. For $\beta>0$, the two bands with $J_x=\pm\frac{3}{2}$ are electron-like (with positive band curvature), whereas the two bands with $J_x=\pm\frac{1}{2}$ are hole-like. Consequently, there must be band-crossing between the $J_x=-\frac{3}{2}$ band and the two $J_x=\pm \frac{1}{2}$ bands for the case with $M>0$. This is illustrated in Fig~\ref{Zeeman}(a). One observes two pairs of band crossing points. The two inner points labeled as $W_1$ are due to crossing between $J_x=-\frac{3}{2}$ and $J_x=-\frac{1}{2}$ bands. They are located at
$
\pm\bm k^{W_1}$, with $\bm k^{W_1}=(\sqrt{M/2\beta},0,0)
$. Meanwhile, the two outer points labeled as $W_2$ are due to crossing between $J_x=-\frac{3}{2}$ and $J_x=\frac{1}{2}$ bands, located at
$
\pm\bm k^{W_2}$ with $\bm k^{W_2}=(\sqrt{M/\beta},0,0)
$.

To characterize the nature of these crossing points, we derive the effective Hamiltonian expanded around these points. First, expanded around $W_1$, we find that it is Weyl point, characterized by
\begin{equation}
H_\text{eff}^{W_1}(\bm q)=v_x q_x\sigma_z+v_y q_y\sigma_y-v_z q_z\sigma_x,
\end{equation}
where $\bm q$ is measured from $\bm k^{W_1}$, $v_x=2\beta k^{W_1}$, $v_y=v_z=\gamma k^{W_1}$, and the Pauli matrices $\sigma_i$ denote the degree of freedom for the two crossing bands.
A Weyl point carries a topological charge of $\pm 1$, which corresponds to the Chern number defined for a constant energy surface enclosing the point. Here, the two Weyl points at $\pm\bm k^{W_1}$ have opposite topological charges, as they are connected by inversion symmetry.

Similarly, we derive the effective model expanded at $W_2$, given by
\begin{eqnarray}
  H_\mathrm{eff}^{W_2}(\bm q)=v q_x\sigma_z+[(\lambda_+ q_+^2+\lambda_-q_-^2)\sigma_+ +{h.c.}],
\end{eqnarray}
where $\bm q$ is measured from $\bm k^{W_2}$, $q_\pm=q_y\pm iq_z$, $\sigma_\pm=\sigma_x\pm i\sigma_y$, $\lambda_\pm=({\sqrt{3}\beta\pm\gamma})/{8}$, and $v =2\beta k^{W_2}$. This Hamiltonian describes a double Weyl point, which has linear dispersion along $q_x$ and quadratic dispersion in the $q_y$-$q_z$ plane. It carries topological charge of $2\ \mathrm{sgn}(|\lambda_+|-|\lambda_-|)=\pm 2$. This pair of double Weyl points are stabilized by the $C_{4x}$ symmetry, as discussed in Ref~\cite{Fang2012}. Perturbations that break $C_{4x}$ would split the double Weyl point into two single Weyl points. For example, consider a perturbation $\delta H=\delta_M\cdot J_y$, which may represent tilting the Zeeman field direction slightly away from the $x$-direction. This breaks the $c_{4x}$ symmetry and destabilize the two double Weyl points. Indeed, in Fig.~\ref{Zeeman}(b), we see that the original double Weyl points no longer exist, and a careful scan shows that they split into two single Weyl points as schematically illustrated in Fig.~\ref{Zeeman}(c). Meanwhile, the two $W_1$ Weyl points are also slightly displaced away from the $k_x$ axis.

Similar analysis can be carried out for Zeeman field along other directions. The result for field along the $[111]$ high-symmetry direction is shown in Fig.~\ref{Zeeman}(d), in which one observes the appearance of two pairs of single Weyl points. Since a Weyl point is topologically robust against weak perturbations, i.e., its topological charge $\pm 1$ cannot be adiabatically removed unless it pair-annihilate with another Weyl point with the opposite charge, these Weyl points should persist as the field direction deviate from the high-symmetry directions. Thus, the topological phase transition from a QCP semimetal to a Weyl semimetal under a Zeeman field should be generic. This observation is consistent with the proposed scenario of creating Weyl points via applied magnetic field in materials such as GdPtBi~\cite{hirschberger2016chiral,cano2017chiral}.

\subsection{Effect of lattice strain}

\begin{figure}
  \includegraphics[width=8.2 cm]{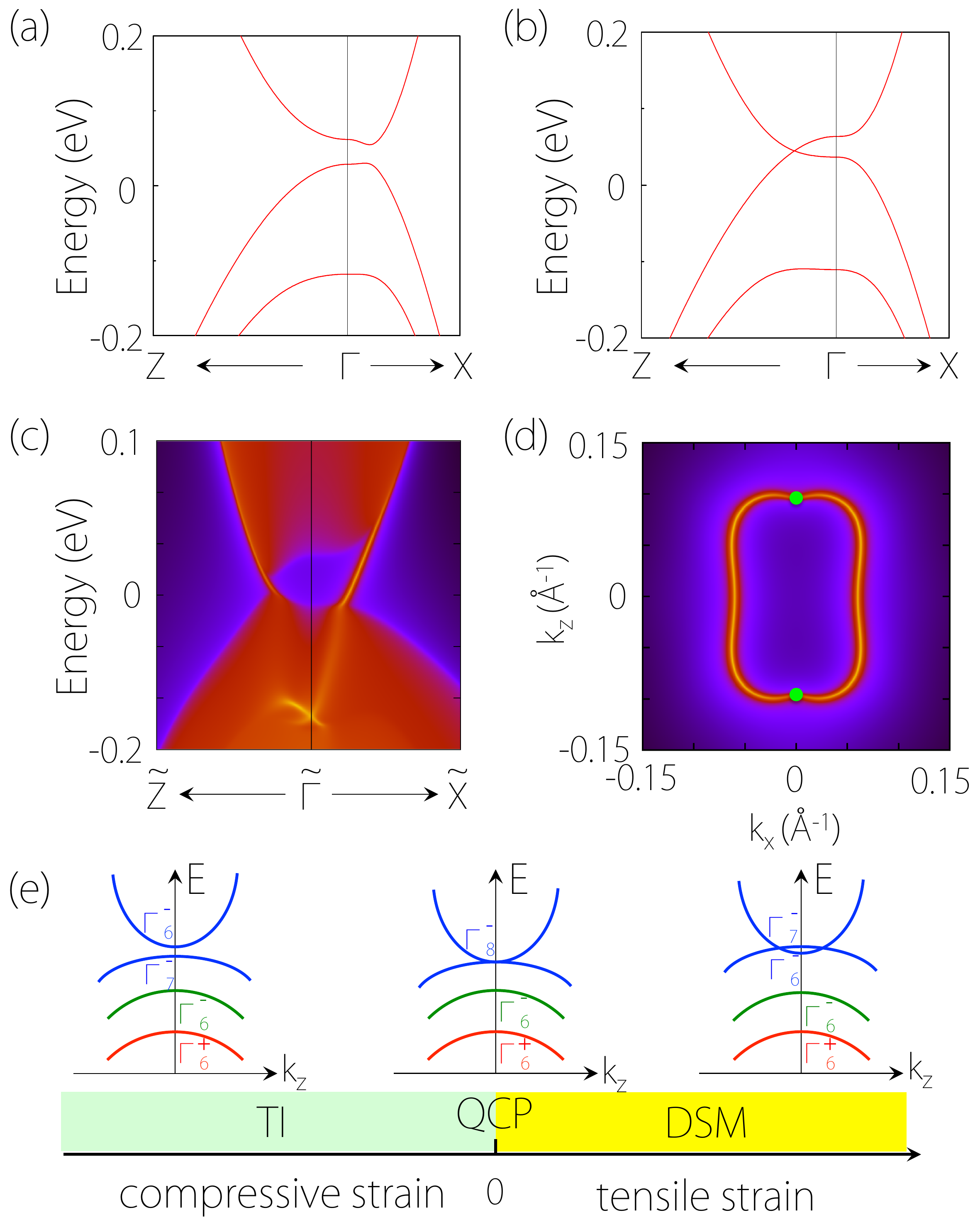}\\
  \caption{Band structures for Cu$_2$Se around the $\Gamma$ point under uniaxial strain along the $[001]$ direction. (a) is for compressive 1\% strain, and (b) is for tensile 1\% strain. (c) Topological surface states on the $(010)$ plane for the topological insulator phase in (a). (d) Fermi arc surface states on the $(101)$ surface for the Dirac semimetal phase in (b), where the projected Dirac nodes are marked as green dots. {(e) Phase diagram for Cu$_2$Se under $[001]$ uniaxial strain in the range of $(-5\%, +5\%)$. The insets above the diagram schematically show the ordering of the low-energy bands around the $\Gamma$ point.} }\label{Cu2Se_strn001}
\end{figure}

Since the QCP point is protected by the crystalline symmetry, it is expected to transform under lattice strains that break the symmetry. We take Cu$_2$Se as example, and use first-principles calculations to investigate the transformation induced by strain.

Let's first consider the strain along the $[001]$ direction, which reduces the symmetry from cubic to tetragonal. The corresponding space group is changed from No.~225 to No.~139. As shown in Fig.~\ref{Cu2Se_strn001}(a), an applied compressive strain of 1\% generates a global bandgap in the band structure. Evaluation of the bulk $\mathbb{Z}_2$ invariant gives a nontrivial $\mathbb{Z}_2=(1;000)$, indicating that the system becomes a strong topological insulator. Physically, this nontrivial band structure arises from the inverted ordering between the Cu-$4s$ states and Cu-$3d$ ($t_{2g}$) states, which we will address in Sec.~VI. Topological insulators feature Dirac-like surface states~\cite{hasan2010colloquium,qi2011topological,shen2012topological}, which we explicitly demonstrate for the current case in Fig.~\ref{Cu2Se_strn001}(c).

A tensile strain along the $[001]$ direction, on the other hand, drives the system towards a Dirac semimetal phase. In Fig.~\ref{Cu2Se_strn001}(b), we show the result for a tensile strain of 1\%. One observes that there are a pair of Dirac points on the $k_z$ axis. Each point is fourfold degenerate, as each crossing band is doubly degenerate due to the time reversal and inversion symmetries. The Dirac points are protected by the $C_{4z}$ symmetry similar to the situation for the Dirac semimetal Cd$_3$As$_2$~\cite{CdAs_wang}. Dirac semimetals often possess Fermi arc surface states~\cite{A3Bi_wang,CdAs_wang}. In Fig.~\ref{Cu2Se_strn001}(d), we plot the surface spectrum and indeed find a pair of surface Fermi arcs connecting the projections of the bulk Dirac points on the side surface.

To summarize the results above, a phase diagram for Cu$_2$Se under  $[001]$ uniaxial strain [in the range of $(-5\%, +5\%)$] is presented in Fig.~\ref{Cu2Se_strn001}(e). One observes that the QCP semimetal state is sitting at the critical point that separates the topological insulator phase and the Dirac semimetal phase.

The results for strains along the $[111]$ direction are qualitatively the same, namely, a compressive strain drives the system towards a topological insulator, whereas a tensile strain drives the system towards a Dirac semimetal. The Dirac points in this case are protected instead by the $C_{3,[111]}$ symmetry similar to the case of Dirac semimetal Na$_3$Bi~\cite{A3Bi_wang}. For strains along other generic directions, Dirac points will generally not be preserved due to the lack of symmetry protection. For example, under strains in the $[110]$ direction, the system becomes topological insulators for both compressive and tensile strains.

\section{Discussion and Conclusion}

In this work, we have revealed the interesting physics associated with QCPs. In the absence of SOC, we can have a triply degenerate QCP stabilized by the point group symmetry. Such point is different from the previously discussed triply degenerate nodal points~\cite{zhu2016triple,winkler2016topological,weng2016topological,weng2016coexistence,chang2017nexus,lv2017observation,zhong2017three,zhang2017coexistence,ma2018three}, which are formed by the crossing between a doubly degenerate band and a nondegenerate band along a three-fold rotational axis. The dispersion is also different. QCPs have quadratic dispersion along all directions, whereas the previously studied triply degenerate nodal points have linear band crossing along the rotational axis.

\begin{figure}[!htp]
{\includegraphics[clip,width=8.4cm]{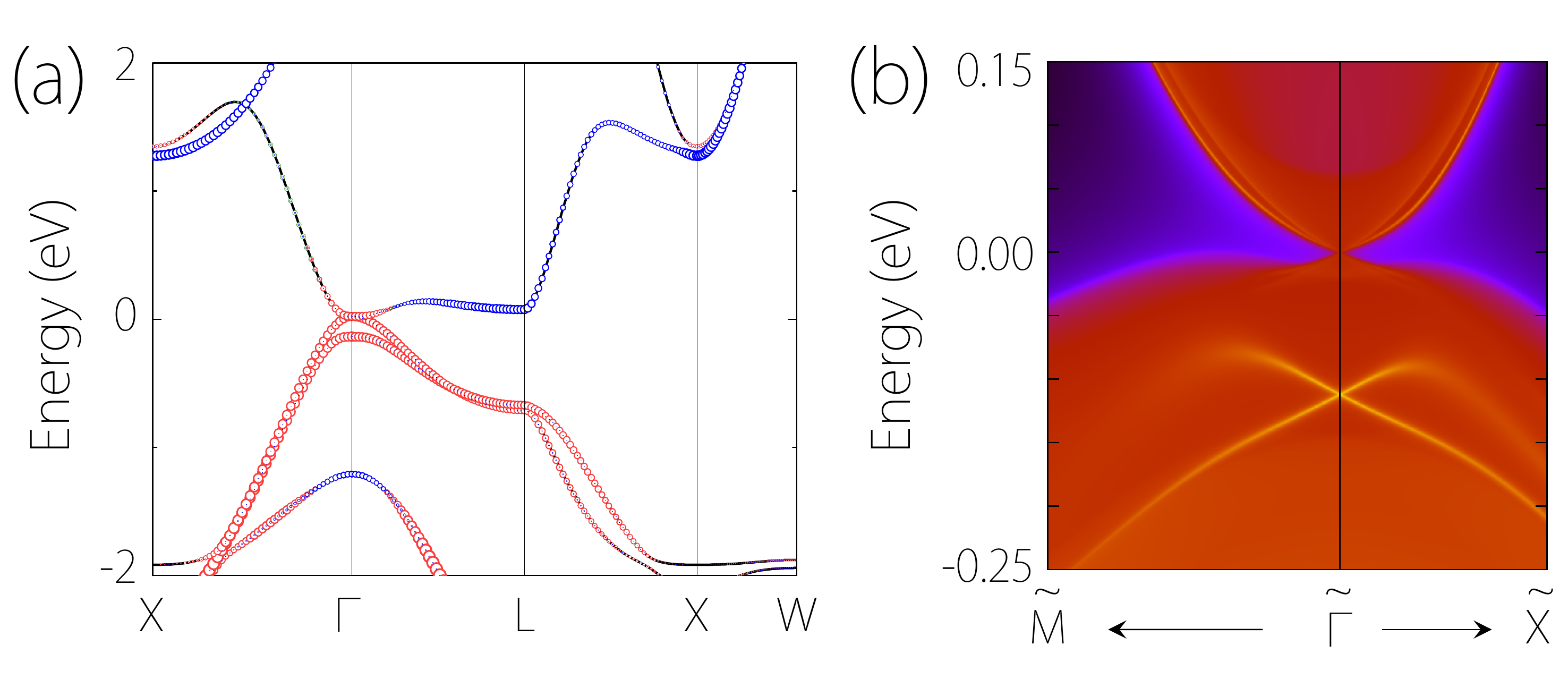}}
\caption{\label{fig:Cu2Se_TSS}
(a) Low-energy band structure for Cu$_2$Se with SOC. The colored circles indicate the weight of Cu-$4s$ (blue) and Cu-$t_{2g}$ (red) orbital characters. (b) Projected spectrum for the (001) surface of Cu$_2$Se, showing Dirac-cone-like topological surface states buried in the projected bulk valence bands.}
\end{figure}

For the QCP semimetal Cu$_2$Se that we discussed here, it is worth noting that the material also has a nontrivial band topology. By analysing the orbital components of the low-energy bands, as in Fig.~\ref{fig:Cu2Se_TSS}(a), we can see that around the $\Gamma$ point, there is prominent Cu-$4s$ component for the valence band around $-1.2$ eV, indicating an inverted band ordering between the Cu-$4s$ states and the Cu-$3d$ ($t_{2g}$) states.
{We find that the band inversion here may also be characterized by a $\mathbb{Z}_2$ invariant similar to that for a topological insulator. This is possible because the two pairs of degenerate states at the QCP share the same parity eigenvalue [which is $-1$, as shown in Fig.~\ref{Cu2Se_strn001}(e)]. Thus, despite the absence of a local gap, the product of parity eigenvalues $\delta_\Gamma$ for the valence states at $\Gamma$ can still be unambiguously defined. Then, we can define the $\mathbb{Z}_2$ invariant $\nu$ via the standard parity analysis at the time-reversal invariant momentum (TRIM) points~\cite{Fu2007}:
\begin{equation}
(-1)^\nu=\prod_i \delta_i,
\end{equation}
where $i$ labels the 8 TRIM points of the BZ. From DFT calculations, we find that $\delta_\Gamma=+1$, $\delta_X=-1$ (at the three $X$ points), and $\delta_L=-1$ (at the four $L$ points). This demonstrates the band inversion at $\Gamma$, and the phase is characterized by a nontrivial $\mathbb{Z}_2$ invariant $\nu=1$.}
Similar to the topological insulators, the band inversion will lead to nontrivial Dirac-cone-like surface states. In the present case, the inversion is between two valence bands, so the surface states appear in an energy range coinciding with the valence bands, as shown in Fig.~\ref{fig:Cu2Se_TSS}(b). This indicates that Cu$_2$Se can be regarded as a \emph{topological QCP semimetal}.

\begin{figure}[!htp]
{\includegraphics[clip,width=8.4cm]{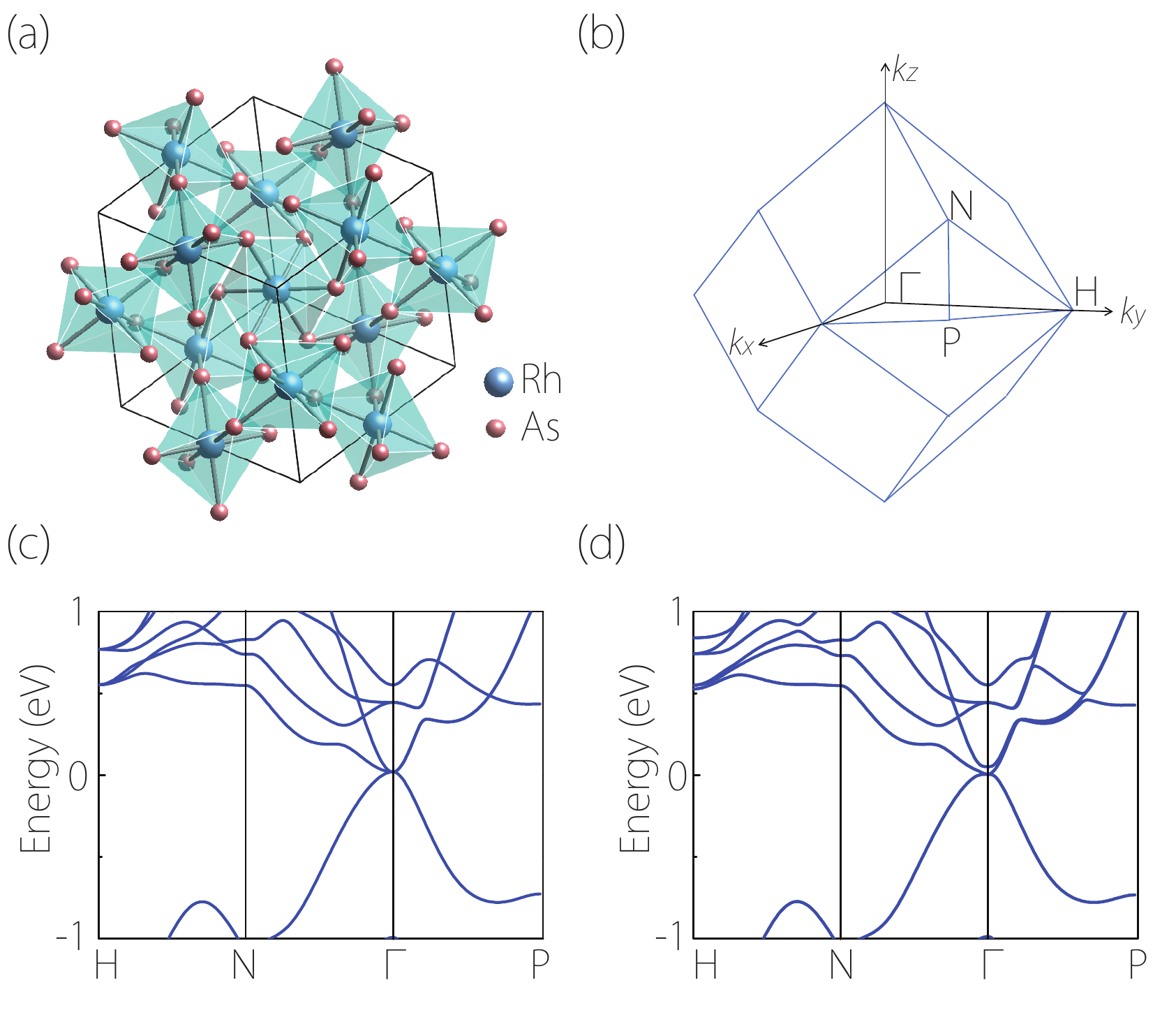}}
\caption{\label{fig:RhAs3}
(a) Crystal structure for RhAs$_3$, showing the primitive cell. (b) Corresponding Brillouin zone. (c,d) Bulk band structures for RhAs$_{3}$ (c) without and (d) with SOC, showing QCPs at the $\Gamma$ point.}
\end{figure}

The presence of QCPs is not unique to the $O_h$ point group discussed here. By inspecting the irreducible representations for other point group symmetries, we find that the QCP may also exist for $O$, $T_d$, and $T_h$ point groups. For example, we consider the material $\rm RhAs_{3}$~\cite{kjekshus1974compounds}, which has a cubic lattice structure with space group No.~204 ($Im\overline{3}$) and with point group $T_h$. It takes the structure as unfilled skutterudite, with four formula units per primitive cell [see Fig.~\ref{fig:RhAs3}(a)]: Rh is at the Wyckoff position $8c$ (0.25000, 0.25000, 0.25000), and As is at $24g$ (0.00000, 0.34843, 0.85345). From the calculated band structures as shown in Figs.~\ref{fig:RhAs3}(c) and \ref{fig:RhAs3}(d), one observes that QCP does appear at the $\Gamma$ point and at Fermi energy. Similar to Cu$_2$Se, it has a triply degenerate QCP in the absence of SOC, and a quadruply degenerate QCP in the presence of SOC, which are protected by the point group symmetry (and also $ \mathcal{T}$).

In conclusion, we have investigated in this work a class of materials which possess quadratic band contact points at the Fermi level. The band dispersion around these points is quadratic along all three directions in $k$-space. We show that in the absence/presence of SOC, there may exist QCPs with threefold/fourfold degeneracy. These points can be protected by the crystalline symmetry. Take the 3D bulk material Cu$_2$Se as an example. We construct effective models to characterize the low-energy fermions near the QCPs. Under strong magnetic field, unlike the usual 3D electron gas, the Landau spectrum for QCP fermions exhibit unconventional features. The QCP semimetal phase is adjacent to a variety of topological phases, such as Weyl semimetal with Weyl/double-Weyl points, topological insulator, and Dirac semimetal. The phase transition can be controlled by Zeeman field or by lattice strain. Our result reveals the QCP semimetal as an interesting phase to study novel emergent fermions, as a parent phase to realize a variety of other topological phases, and as an intriguing platform to study topological phase transitions.

\emph{Note added.} After submission of this work, we became aware of Refs.~\cite{Ghorashi2017,Zhang2018,Ghorashi2018}, which studied the quadruply-degenerate QCP state in $\alpha$-Sn~\cite{Zhang2018}, and its properties under light irradiation~\cite{Ghorashi2018} and disorder scattering~\cite{Ghorashi2017}.

\begin{acknowledgements}
The authors thank Quansheng Wu, Jinqi Wu, and D. L. Deng for valuable discussions. This work was supported by the Singapore Ministry of Education AcRF Tier 2 (Grant No.~MOE2015-T2-2-144). Z. Zhu was supported by the National Natural Science Foundation of China (NSFC) (No.~11704117) and the start-up funds from Hunan Normal University. X.-L. Sheng was supported by NSFC (No.~11504013). We acknowledge computational support from the Texas Advanced Computing Center, the National Supercomputing Centre Singapore (https://www.nscc.sg), and H2 clusters in Xi'an Jiaotong University.
\end{acknowledgements}

\begin{appendix}

\renewcommand{\theequation}{A\arabic{equation}}
\setcounter{equation}{0}
\renewcommand{\thefigure}{A\arabic{figure}}
\setcounter{figure}{0}
\renewcommand{\thetable}{A\arabic{table}}
\setcounter{table}{0}

\begin{figure}[t!]
\includegraphics[width=8cm]{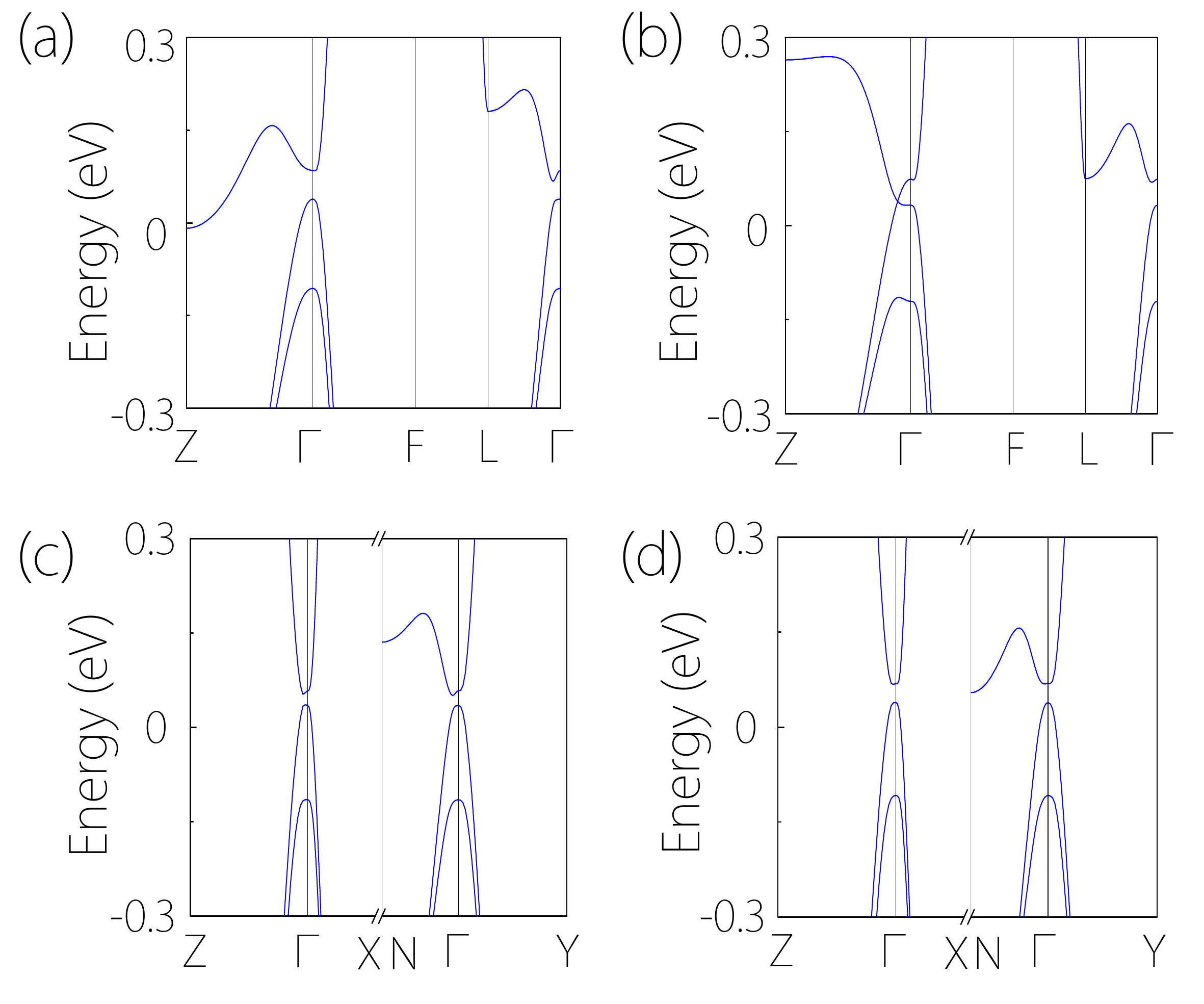}
\caption{(a,b) Band structures for Cu$_2$Se under uniaxial strain along the $[111]$ direction. Here, the angle between the crystal axis ($a$, $b$ and $c$) is changed from 90$^{\circ}$ to (a) 91$^{\circ}$ and (b) 89$^{\circ}$, respectively. (c,d) Band structures for Cu$_2$Se under uniaxial (c) compressive and (d) tensile strains of 1\% along the $[110]$ direction.}
\label{Cu2Se_strn111}
\end{figure}

\section{First-principle calculation}
\label{sec:meth}
The first-principles calculations were performed using the Vienna \emph{ab initio} simulation package~\cite{PhysRevB.54.11169,kresse1999ultrasoft} with the projector augmented wave method~\cite{PhysRevB.50.17953}. For the exchange correlation energy, the generalized gradient approximation (GGA) with the Perdew-Burke-Ernzerhof (PBE) realization~\cite{PhysRevLett.77.3865} was adopted. The energy cutoff was set to 460 eV, and a $11\times11\times11$ Monkhorst-Pack mesh is used for the Brillouin zone sampling. For the electronic self-consistent calculations, the convergence criterion for the total energy was set to be $10^{-8}$ eV. The lattice constants are fully relaxed until the total energy is converged with a tolerance less than $10^{-7}$ eV and the residual forces on atoms are below $10^{-3}$ eV/{\AA}. The topological invariants are evaluated using the Z2pack code~\cite{gresch2017z2pack}.
The surface states are investigated using the iterative Green's function method~\cite{sancho1985highly} as implemented in the WannierTools package~\cite{wu2017wanniertools}.

\section{Results for strain in the [111] and [110] direction}

Under an applied strain along the $[111]$ direction, the space group of Cu$_2$Se is changed from No.~225 to No.~166. The band structures with SOC are shown in Fig.~\ref{Cu2Se_strn111}(a) and \ref{Cu2Se_strn111}(b) for compressive and tensile strains, where the angle between the crystal axis ($a$, $b$ and $c$) is changed from 90$^{\circ}$ to 91$^{\circ}$ and 89$^{\circ}$, respectively. Similar to HgTe, a compressive strain drives the system towards a topological insulator phase [see Fig.~\ref{Cu2Se_strn111}(a), but here the global gap closes indirectly, making the system a topological metal], whereas a tensile strain drives the system towards a Dirac semimetal phase [see Fig.~\ref{Cu2Se_strn111}(b)]. On the other hand, under strains along the $[110]$ direction,  the space group of Cu$_2$Se is changed into No.~71. The system becomes topological insulators for both compressive and tensile strains [see Fig.~\ref{Cu2Se_strn111}(c) and \ref{Cu2Se_strn111}(d)].

\end{appendix}

\bibliography{QCPSM_refs}

\end{document}